\documentclass{iau}

\usepackage{amsmath}
\usepackage{graphicx}
\usepackage{multirow}

\newcommand{\Ms}{\ensuremath{\rm M_{\odot}}}

\newcommand{\Mzam}{\ensuremath{ M_{\rm ZAMS}}}

\begin{document}

\lefttitle{Cristiano Ugolini}
\righttitle{Piling up in the darkness: Features of the BBH mass distribution from isolated binaries}

\jnlPage{xxx}{xxx}
\jnlDoiYr{xxx}
\doival{xxxxx}

\aopheadtitle{IAU Symposium 398/MODEST-25}
\editors{xxx}
\title{Piling up in the darkness: Features of the BBH mass distribution from isolated binaries}

\author{Cristiano Ugolini}
\affiliation{Gran Sasso Science Institute (GSSI),                L'Aquila (Italy), Viale Francesco                  Crispi 7\\
            INFN, Laboratori Nazionali del Gran Sasso, 67100 Assergi, Italy\\}        

\begin{abstract}
After the third LIGO--Virgo--KAGRA observing run, the number of detected binary black hole (BBH) mergers became sufficient to identify statistical features of the population. 
We explore how different prescriptions for the final fate of massive stars and key binary-evolution processes shape isolated binaries and their remnants. 
Using \textsc{sevn}, we evolved $10^{7}$ binaries across 15 metallicities, 3 core-collapse supernova models, 4 PPISN models, and 6 common-envelope (CE) prescriptions, for a total of 990 runs ($9.9 \times 10^{9}$ systems). 
Both single- and binary-star physics shape the BH mass distribution: single-star processes control the high-mass tail ($M_{\rm BH} \geq 45\,M_{\odot}$), while binary evolution produces pile-ups in specific intervals. 
In particular, the bump at $\sim 35\,M_{\odot}$, often attributed to PPISNe, also emerges from binaries evolving only through stable mass transfer, without CE. 
Finally, we test a top-heavy IMF, finding it boosts merger numbers and alters the abundance of systems with a given primary BH mass.
\end{abstract}


\maketitle

\section{Introduction}

The three observing runs of the LIGO–Virgo–KAGRA (LVK; \citealt{ligoscientificcollaborationandvirgocollaborationGW190521BinaryBlack2020,LVK_2021,LVK_2023}) collaboration revealed a population of stellar black holes (BHs) more massive than those known from X-ray binaries \citep[][]{MillerJones2021}. By GWTC-3, 90 BBH mergers\footnote{O4 is expected to yield at least \href{https://www.ligo.caltech.edu/news/ligo20250320}{200 events}} enabled inference of the BH mass spectrum, favoring a \textsc{Power Law + Peak} model with a Gaussian excess at $M_{\rm BH}\sim34\,\Ms$ \citep{LVK_2023,callisterParameterFreeTourBinary2023}. This feature may trace specific evolutionary pathways \citep{LVK_2021}.

Theoretical models of BBH formation remain uncertain \citep[see review by][]{Spera2022}. Two main channels are proposed: (i) isolated binaries \citep[e.g.][]{mapelliMassiveBlackHole2016a,speraMergingBlackHole2019a,iorioCompactObjectMergers2023} or (ii) dynamical assembly in dense clusters \citep[e.g.][]{Rodriguez2015,Mapelli2016, Arca_Sedda_2023}. 
Both are affected by uncertainties in winds, rotation, and SN physics \citep[e.g.][]{burrowsThreedimensionalSupernovaExplosion2019,Ugolini2025}.  
Pair-instability processes, dividing into PISNe and PPISNe \citep{fowlerNeutrinoProcessesPair1964}, shape the upper BH mass limit \citep{woosleyPulsationalPairinstabilitySupernovae2017}. 
While the $35\,\Ms$ bump has been attributed to PPISNe \citep{belczynskiEffectPairinstabilityMass2016}, recent works emphasize binary interactions such as SMT and CE \citep{vansonNoPeaksValleys2022a,dorozsmaiImportanceStableMass2024b}.

Here we explore how single-star and binary evolution combine to shape the BH mass spectrum using \textsc{SEVN} \citep{speraMergingBlackHole2019a,iorioCompactObjectMergers2023}. 
Section~\ref{sec: Methods} summarizes the setup, Section~\ref{sec: Results} presents the impact of CE efficiency, and Section~\ref{sec: Discussion} discusses IMF effects.

\section{Methods}
\label{sec: Methods}

We employ the population-synthesis code \textsc{SEVN} \citep{speraMassSpectrumCompact2015a}, which interpolates stellar tracks from PARSEC v2.7.0 \citep{bressanPARSECStellarTracks2012a}. 
Initial conditions follow \cite{iorioCompactObjectMergers2023}: primary masses are drawn from a Kroupa IMF in $[8,150]\,\Ms$, and secondaries from $[5,150]\,\Ms$. 
This analysis focuses on a subsample of the simulations presented in \cite{Ugolini_bump_2025}, consisting of $10^{7}$ binaries evolved for each of 15 metallicities and 6 CE prescriptions, plus an additional run adopting a Larson IMF \citep{Larson1998}.

\section{Effect of BSE on the \texorpdfstring{$35\,\Ms$}{35 Ms} peak}
\label{sec: Results}

\begin{figure*}[htb]
    \centering
    \includegraphics[width=\textwidth]{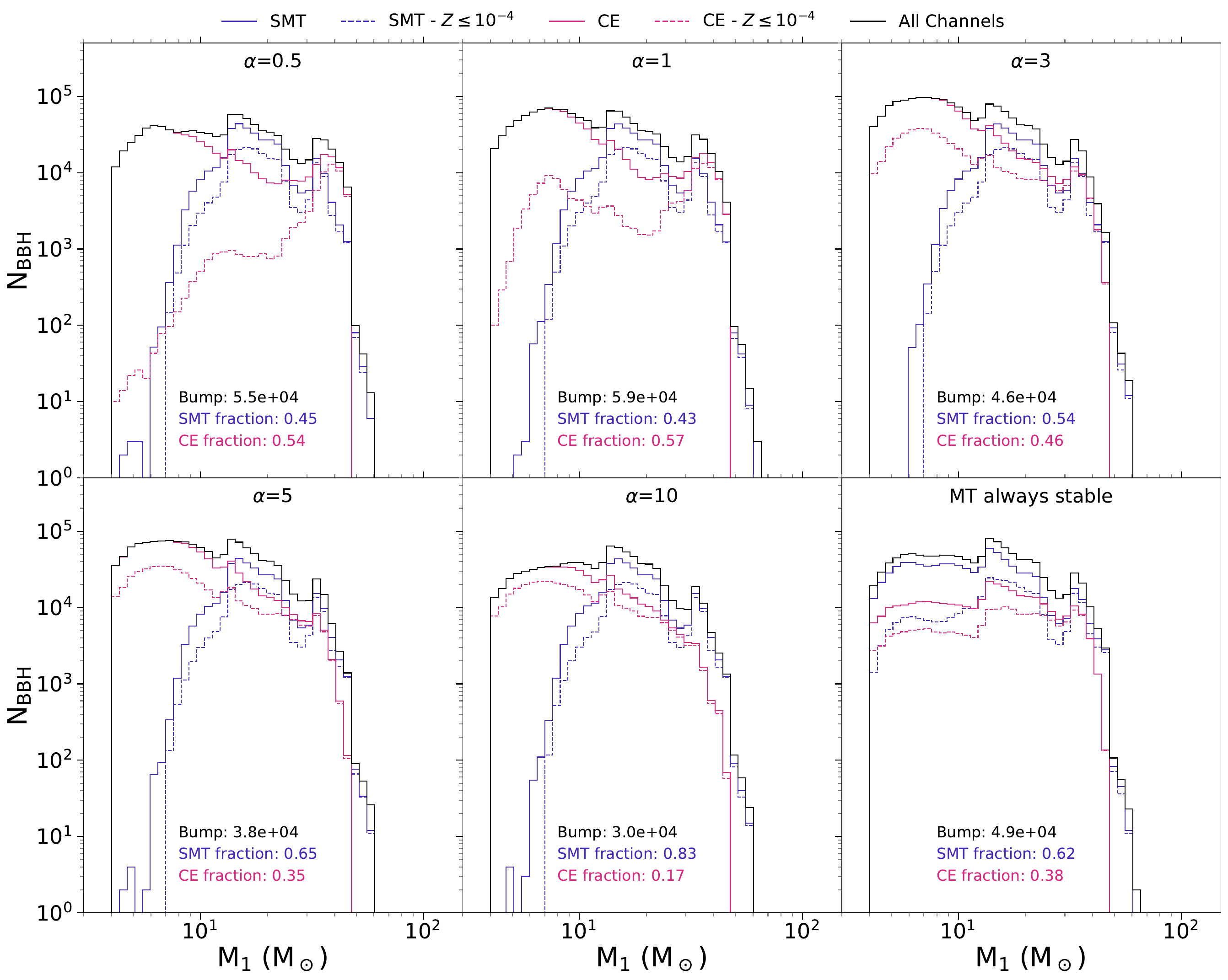}
    \caption{Primary BH mass distribution of merging BBHs for different CE efficiencies $\alpha$.
    Upper row: $\alpha=0.5,1,3$; lower row: $\alpha=5,10$, and the always-stable MT case.
    Blue: SMT channel; purple: CE channel. Dashed: metal-poor (\(Z\leq 5\times10^{-4}\)); solid: full population.
    The black curve is the total over all the channels.}
    \label{fig: BSE bump - alpha variation}
\end{figure*}

We assess how the CE efficiency parameter $\alpha$ shape the BH mass spectrum under our fiducial ccSN+PPISN combination (\texttt{delayed\_Gauss}+{\tt M20}). We also consider an “always-stable MT” setup, where CE can still occur only if both stars overflow their Roche lobes (treated with $\alpha=3$ when it happens).

\noindent\textbf{CE parameter $\alpha=0.5$.} CE ejection is inefficient, so many systems merge as stars rather than form BBHs. Among survivors, the long-lasting CE phase shrinks orbits efficiently, making CE the dominant channel in the BBH sample. Indeed, the CE channel supplies $\sim54\%$ of BHs in the $32$–$37\,\Ms$ bump, with CE and SMT nearly degenerate at $32$–$34.5\,\Ms$, and CE producing more BHs at $35$–$37\,\Ms$. SMT builds up BHs at $\sim13$–$22\,\Ms$ (about $65\%$ of SMT-formed primaries).

\noindent\textbf{CE parameter $\alpha=1$.} CE shrinks orbits less than at $\alpha=0.5$, thus more binaries successfully eject the envelope before merging, increasing the overall merger yield. CE contributes to $57\%$ of the BBH in the bump. As at $\alpha=0.5$, channels are degenerate at $32$–$34.5\,\Ms$; CE contributes more at $35$–$37.5\,\Ms$.

\noindent\textbf{CE parameter $\alpha=3$.} Maximum amount of BBH among all the setups. The $32$–$37\,\Ms$ bump is now primarily from SMT ($\sim54\%$), with CE and SMT contributing similarly at $35$–$37\,\Ms$.

\noindent\textbf{CE parameter $\alpha=5$.} 
CE still dominates for $M_{\rm BH}\leq 14\,\Ms$, while SMT dominates at $16$–$22\,\Ms$ and in the $32$–$37\,\Ms$ bump. In this bump, $65\%$ of systems evolved without CE (pure SMT). CE systems contributing to the bump decline because prompt ejection leaves separations too wide to merge within a Hubble time.

\noindent\textbf{CE parameter $\alpha=10$.} SMT shapes the higher-mass distribution. In the $32-37\,\Ms$ mass-interval, $83\%$ of merging systems never undergo CE. The total number of mergers and the general shape resemble $\alpha=0.5$, but the bump is narrower for high $\alpha$ (concentrated at $32$–$34.5\,\Ms$) and broader for low $\alpha$ (extending to $\sim37\,\Ms$).

\noindent\textbf{Always-stable MT.} CE becomes secondary (it can still occur if both stars overflow their RLs). Even the least-massive BHs are then mostly formed via SMT, yet CE still contributes $\sim38\%$ of primaries in the bump.

\medskip
\noindent\textbf{Global feature.} Across models, primaries tend to accumulate at $32$–$37\,\Ms$: about $7\%$ of BHs form in this interval, versus only $\sim3\%$ and $\sim2\%$ in the adjacent $27$–$32\,\Ms$ and $37$–$42\,\Ms$ bins, respectively. In sum, CE predominantly seeds the low-mass end, while SMT governs the $32$–$37\,\Ms$ excess, with their balance regulated by $\alpha$.

\section{Effect of a top-heavy IMF}
\label{sec: Discussion}

\begin{figure*}[htb]
    \centering
    \includegraphics[width=\textwidth]{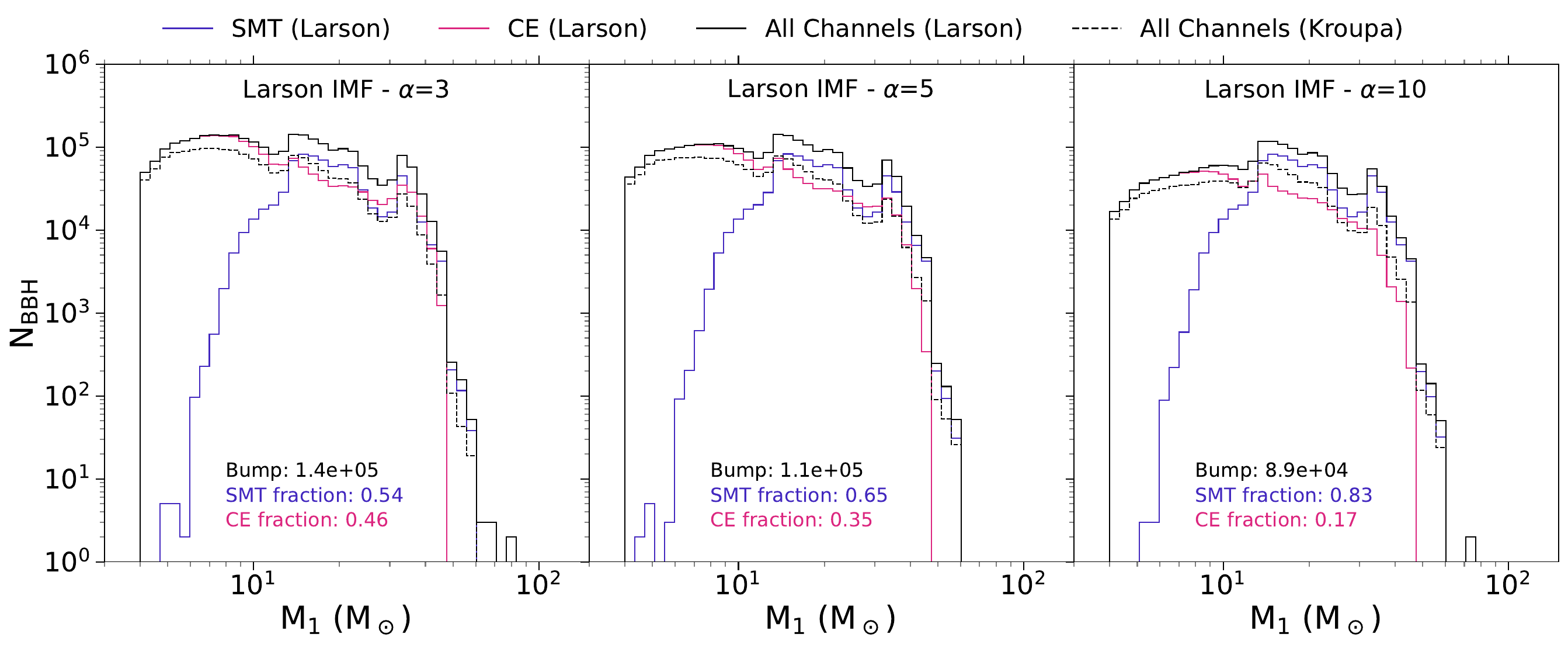}
    \caption{Primary BH mass distribution of merging BBHs for a Larson IMF and different CE efficiencies $\alpha=3,5,10$.
    Blue: SMT; purple: CE. Black solid: Larson IMF; black dashed: Kroupa IMF.}
    \label{fig: Bump - IMF variation}
\end{figure*}

Stars that populate the $32$–$37\,\Ms$ bump typically have $\Mzam \gtrsim 70\,\Ms$, making the IMF a key factor in setting their relative abundance. To test this, we adopt a Larson IMF \citep{Larson1998}, usually invoked for Pop~III stars but applied here across all metallicities to probe the effect of favoring massive stars. 
Other initial conditions follow our fiducial setup (see Section~\ref{sec: Methods}). Figure~\ref{fig: Bump - IMF variation} shows the resulting distributions for $\alpha=3,5,10$.  

\noindent\textbf{CE parameter $\alpha=3$.} As in the Kroupa IMF case, SMT contributes $54\%$ of bump primaries and CE $46\%$. The Larson IMF, however, boosts the overall merger yield by $\sim67\%$ and nearly triples the number of systems in the $32$–$37\,\Ms$ range. The abundance of low-mass BHs ($M_{\rm BH}\lesssim 10\,\Ms$) is only mildly affected.  

\noindent\textbf{CE parameter $\alpha=5$.} The bump is dominated by SMT ($65\%$) with CE contributing $35\%$, similar to the Kroupa case, but the absolute number of bump BHs again rises by a factor $\sim3$, and the total mergers increase by $\sim68\%$.  

\noindent\textbf{CE parameter $\alpha=10$.} As in the Kroupa IMF case, $83\%$ of bump BHs evolve without CE. The Larson IMF triples the bump population relative to Kroupa and raises the overall merger count by $\sim80\%$.  

\medskip
In summary, adopting a top-heavy IMF greatly amplifies the $32$–$37\,\Ms$ feature and the total merger rate, while leaving the qualitative role of CE vs. SMT largely unchanged.







\section{Summary and conclusions} 
\label{sec: summary and conclusions}

Large-scale population synthesis with \textsc{SEVN} shows that:
\begin{itemize}
\item The $32$–$37\,\Ms$ bump arises from a combination of CE and SMT pathways. CE dominates for $\alpha\le1$, SMT for $\alpha\ge3$.
\item A top-heavy IMF amplifies the bump and merger rate without changing the CE/SMT balance.
\item CE efficiency and IMF leave measurable signatures on the BH mass spectrum, offering GW observations a way to constrain them.
\end{itemize}
Future work will integrate these results into the semi-analytical BPOP model \citep{Arca_Sedda_2023} to include metallicity and star-formation histories.

\begin{acknowledgements}
The author acknowledges Mario Spera, Marco Limongi, and Manuel Arca sedda for useful discussion and comments
\end{acknowledgements}

\bibliographystyle{mnras}
\bibliography{Sample}

\end{document}